\documentstyle[12pt,a4]{article}
\begin{document}

\setlength{\baselineskip}{24pt}

\begin{center}
\today\\[1cm]{\large {\bf The Generalized Gutzwiller Method for $n\geq 2$
Correlated Orbitals: Itinerant Ferromagnetism in \\}}${\bf d}${\large {\bf ($%
e_g)-$ bands}\\[1cm]}{\normalsize {J. B\"{u}nemann and W. Weber \\Institut
f\"{u}r Physik, Universit\"{a}t Dortmund, \\D-44221 Dortmund, Germany}\\[1cm]%
}
\end{center}

{\bf Abstract}:\\Using the generalized Gutzwiller method we present results
on the ferromagnetic behavior of extended Hubbard models with two degenerate 
$d$($e_g$) orbitals. We find significant differences to results obtained
from Hartree-Fock theory.\\[5mm]{\bf Keywords}: \\correlated electrons,
extended Hubbard models, Gutzwiller method, itinerant ferromagnetism\\{\bf %
Postal Address}: \\Prof. Dr. Werner Weber\\Universit\"{a}t Dortmund,
Institut f\"{u}r Physik,\\Lehrstuhl f\"{u}r Theoretische Physik II\\44221
Dortmund\\{\bf Fax}:\\++49 0231 755 3569\\{\bf E-mail}: \\%
weber@fkt.physik.uni-dortmund.de \newpage
Recently, the Gutzwiller variational method has been generalized for $n\geq
2 $ correlated orbitals per site \cite{c3}. In this paper we present studies
of the itinerant ferromagnetism using a two-band Hubbard model with
degenerate orbitals and general on-site interactions.

The one-band Gutzwiller variational wavefunction is given by \cite{c1} 
\begin{equation}
\left| \Psi \right\rangle \equiv g^{\widehat{D}}\left| \Psi _0\right\rangle
=\prod_{s=1}^L\left[ 1-(1-g)\widehat{D}_s\right] \left| \Psi _0\right\rangle
\;,  \label{1}
\end{equation}
where $\left| \Psi _0\right\rangle $ is an uncorrelated wavefunction on a
lattice of $L$ sites and the operators $\widehat{D}_s\equiv \widehat{n}%
_{s\uparrow }\widehat{n}_{s\downarrow }$ measure the double occupancy of
sites $s$. By combinatorics, one is led to the hopping 'loss factors' 
\begin{equation}
q_k\equiv \frac 1{m_k(1-m_k)}\left[ \sqrt{\overline{m}_k\overline{m}_0}+%
\sqrt{\overline{m}_{kl}\overline{m}_l}\,\right] ^2\;,  \label{2}
\end{equation}
where $m_0,\,m_1,\,m_2,\,m_{12}\;(\overline{m}_0,\,\overline{m}_1,\,%
\overline{m}_2,\overline{m}_{12})$ are the respective gross (net)
occupancies of sites: empty, spin-up, spin-down, double. There are relations
between gross and net occupancies such as $\overline{m}_1=m_1-\overline{m}%
_{12}$ or $m_0=\overline{m}_0=1-\overline{m}_1-\overline{m}_2-\overline{m}%
_{12}$.

The extension to cases of $n\geq 2$ correlated orbitals ($a,b,..$) leads to $%
J_n=4^n$ different possibilities for a single site occupancy, and $%
K_n=J_n-(2n+1)$ of them represent multiple ones. The basic idea of our
generalization was to include all $K_n$ multiple occupancy operators.

Using the notation $a\!\!\uparrow \!\!\widehat{=}1,$ $a\!\!\downarrow \!\!%
\widehat{=}2,$ $b\!\!\uparrow \!\!\widehat{=}3,$ $b\!\!\downarrow \!\!%
\widehat{=}4,...,$ the resulting generalized hopping loss factors have been
found to be 
\begin{eqnarray}
&&q_{kk} =\frac 1{m_k(1-m_k)}[\sqrt{\overline{m}_k\overline{m}_0}%
+\sum_l{}^{^{\prime }}\sqrt{\overline{m}_{kl}\overline{m}_l}  \label{6b} \\
&&+\sum_{l,p}{}^{^{\prime \prime }}\sqrt{\overline{m}_{klp}\overline{m}_{lp}}%
+\sum_{l,p,q}{}^{^{\prime \prime \prime }}\sqrt{\overline{m}_{klpq}\overline{%
m}_{lpq}}+...]^2  \nonumber \\
&&q_{kl}^2=q_{kk}q_{ll}\;.  \label{6}
\end{eqnarray}
It is now possible to investigate extensions of the Hubbard model for
arbitrary numbers of orbitals $\alpha ,\beta $ and more general on-site
interactions: 
\begin{eqnarray}
\widehat{H} &=&\sum_{\alpha ,\beta ,s,t,\sigma }T_{st}^{\alpha \beta }%
\widehat{\alpha }_{s\sigma }^{+}\widehat{\beta }_{t\sigma }+\sum_{\alpha
,\beta ,s,\sigma ,\sigma ^{\prime }}\!\!\!\!^{^{\prime }}U^{\alpha \beta }%
\widehat{n}_{s\sigma }^\alpha \widehat{n}_{s\sigma ^{\prime }}^\beta 
\nonumber \\
&&+\sum_{\alpha ,\beta ,s,\sigma ,\sigma ^{\prime }}\!\!\!\!^{^{\prime
}}J^{\alpha \beta }\widehat{\alpha }_{s\sigma }^{+}\widehat{\beta }_{s\sigma
^{\prime }}^{+}\widehat{\alpha }_{s\sigma ^{\prime }}\widehat{\beta }%
_{s\sigma }\;.  \label{7}
\end{eqnarray}

For our present study we have focussed on a ferromagnetic case; i.e. we have
chosen $N$-particle wave functions $\left| \Psi \right\rangle ,\,\left| \Psi
_0\right\rangle $ which allow that $m_1=m_3=m_{+}=\left\langle \widehat{n}%
_{\uparrow }\right\rangle \neq \left\langle n_{\downarrow }\right\rangle
=m_{-}$. This spin-splitting is controlled by an additional variational
parameter $\Delta $ ($m_{+}=m+\Delta ;\,m_{-}=m-\Delta $), representing the
magnetization. We have studied the $n=2$ Hamiltonian with two $e_g-$type
orbitals on a simple cubic lattice and have included first nearest neighbor $%
\left( 1NN\right) $ and $2NN$ hopping terms given by \cite{c2b}: $%
T_{dd\sigma }(1NN)=1eV,$ $T_{dd\sigma }(2NN)=0.25eV$, and $T_{dd\delta
}:T_{dd\pi }:T_{dd\sigma }=0.1:-0.3:1$. Further, there are three interaction
parameters $U^{\alpha \alpha }\equiv U,\;U^{\alpha \beta }\equiv U^{\prime }$
and $J^{\alpha \beta }\equiv J$. We employ the relation $2J=U-U^{\prime }$,
which is valid for $e_g-$ orbitals in the limit of vanishing configuration
interaction \cite{c2}.

There are seven variational parameters representing the multiple
occupancies: $f\equiv \overline{m}_{1234},\;t_{+}\equiv \overline{m}_{123}=%
\overline{m}_{134},\;t_{-}\equiv \overline{m}_{124}=\overline{m}%
_{234},\;d_{t+}\equiv \overline{m}_{13},\;d_{-}\equiv \overline{m}%
_{24}\;d_d\equiv \overline{m}_{12}=\overline{m}_{34},\;d_o\equiv \overline{m}%
_{14}=\overline{m}_{23}$. Further, there are two different loss factors $%
q_{+}\equiv q_{11}=q_{33}=q_{13}=q_{31}$ and $q_{-}\equiv
q_{22}=q_{44}=q_{24}=q_{42}$, with 
\begin{eqnarray}
q_\mu  &=&\frac 1{m_\mu (1-m_\mu )}[\sqrt{\overline{m}_\mu }(\sqrt{\overline{%
m}_0}+\sqrt{d_{t\mu }})  \label{9} \\
&&+(\sqrt{d_d}+\sqrt{d_o})(\sqrt{\overline{m}_{(-\mu )}}+\sqrt{t_\mu }) \nonumber\\
&&+\sqrt{t_{(-\mu )}}(\sqrt{f}+\sqrt{d_{t(-\mu )}})]^2\,.  \nonumber
\label{30}
\end{eqnarray}
The ground state energy function is 
\begin{eqnarray}
&&E=q_{+}\overline{\epsilon }(m_{+})+q_{-}\overline{\epsilon }(m_{-})
\label{10} \\
&&+2Ud_d+2U^{\prime }d_o+(U^{\prime }-J)(d_{t+}+d_{t-})+  \nonumber \\
&&+(2U+4U^{\prime }-2J)(t_{+}+t_{-}+f)\;,  \nonumber  \label{31}
\end{eqnarray}
with $\overline{\epsilon }(m)$ being the kinetic energy of the uncorrelated
bands. Note that the number of electrons per atom is given by $e/a=4m$. We
have studied band fillings, which, for the paramagnetic case, are close to
the biggest peak in the single particle DOS.

Fig. 1 shows the magnetization $\Delta $ as a function of $U$ (with $%
J=0.2\,U $) for the values $m=0.35$ (Fig 1a) and $m=0.3$ (Fig. 1b). Also
shown are the values of $N(E_F)$. When $e/a$ is chosen so that $N(E_F)$ of
the uncorrelated case coincides with the DOS peak ($m=0.3$), there exists a
kind of Stoner criterion for the ferromagnetic instability. Then, $\Delta $
starts as a continuous function of ($U,U^{\prime },J$) (Fig. 1b). Away from
this peak, there is a first order transition to a state of finite
magnetization before the Stoner criterion is met (Fig. 1a). We observe
further discontinuous increases of the magnetization with increasing
interaction ($U,U^{\prime },J$). These jumps are related to structures in
the DOS\ which pass through the Fermi level with increasing splitting of
majority and minority bands (see Fig. 1a).

A Hartree-Fock treatment of the model leads to onsets of ferromagnetism at
much smaller values of the interaction parameters $U,U^{\prime },J$ (Fig.
2). The first order transition from $\Delta =0$ to finite $\Delta $ is also
seen in HF theory; however no further jumps in magnetization are observed.

The magnetic phase diagram (Fig. 3) elucidates the role of the on-site
exchange for the formation of itinerant ferromagnetism.

In conclusion, our study of itinerant ferromagnetism indicates significant
differences between the Gutzwiller method and the Hartree-Fock theory. We
hope that the generalized Gutzwiller method will enable us to judge the
quality of effective one-particle theories for itinerant magnetism such as
spin density functional theory for 3d transition metals.

Acknowledgments: This work has been partly supported by the European Union
Human Capital and Mobility program, Project No. CHRX-CT 93-0332 
% now the references. delete or change fake bibitem. delete next three
%   lines and directly read in your .bbl file if you use bibtex.

% \begin{figure}
% \caption{}
% \label{}
% \end{figure}

%  fig. 1
\newpage
{\large {\bf Figures} }{\normalsize \\[1cm]{Figure 1: Fermi density of
states }}$N_{EF}${\normalsize {\ and interaction }}$U${\normalsize {\ (with }%
}$J=0.2\,U${\normalsize {) versus magnetization }}$\Delta ${\normalsize {\
for two bandfillings }}$m${\normalsize {. The dotted vertical lines in (a)
indicate first order changes in }}$\Delta ${\normalsize {.}\\[0.5cm]{Figure
2: Comparison of Gutzwiller and Hartree-Fock results for $\Delta $ versus $U$
($J=0.2\,U$). Dashed lines: $m=0.3$, solid lines: $m=0.35$. Inset shows the
jump in $\Delta $ near $U=7.95$ in the Gutzwiller result for $m=0.35$.}%
\\[0.5cm]{Figure 3: Phase diagram in the $J/U$ versus $U$ plane for the
Gutzwiller results with $m=0.3$ (note that $U^{\prime }=U-2J$).} }

{\normalsize % tables follow here
% \begin{table}
% \caption{}
% \label{}
% \end{tabular}
% \end{table}
}

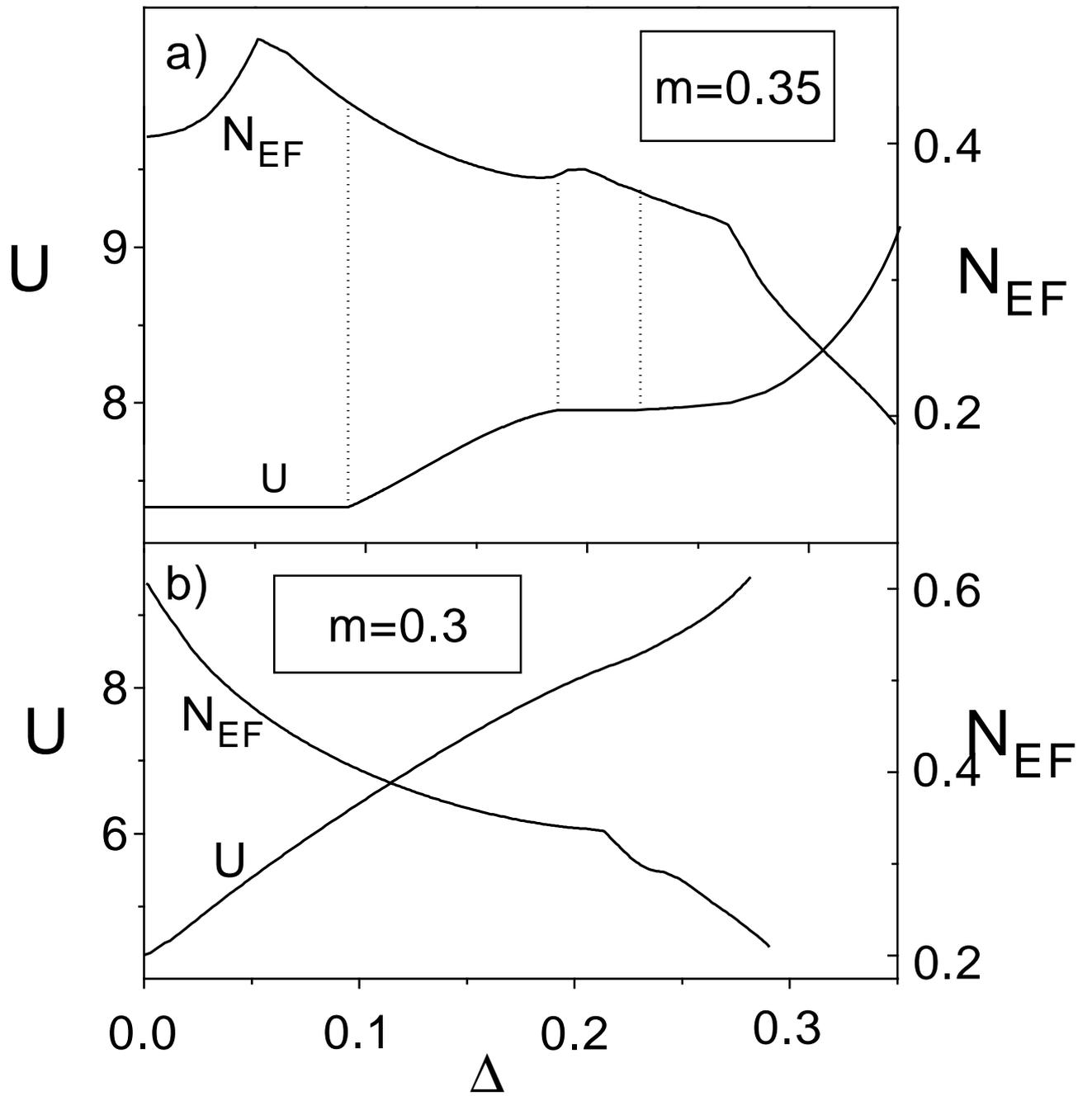
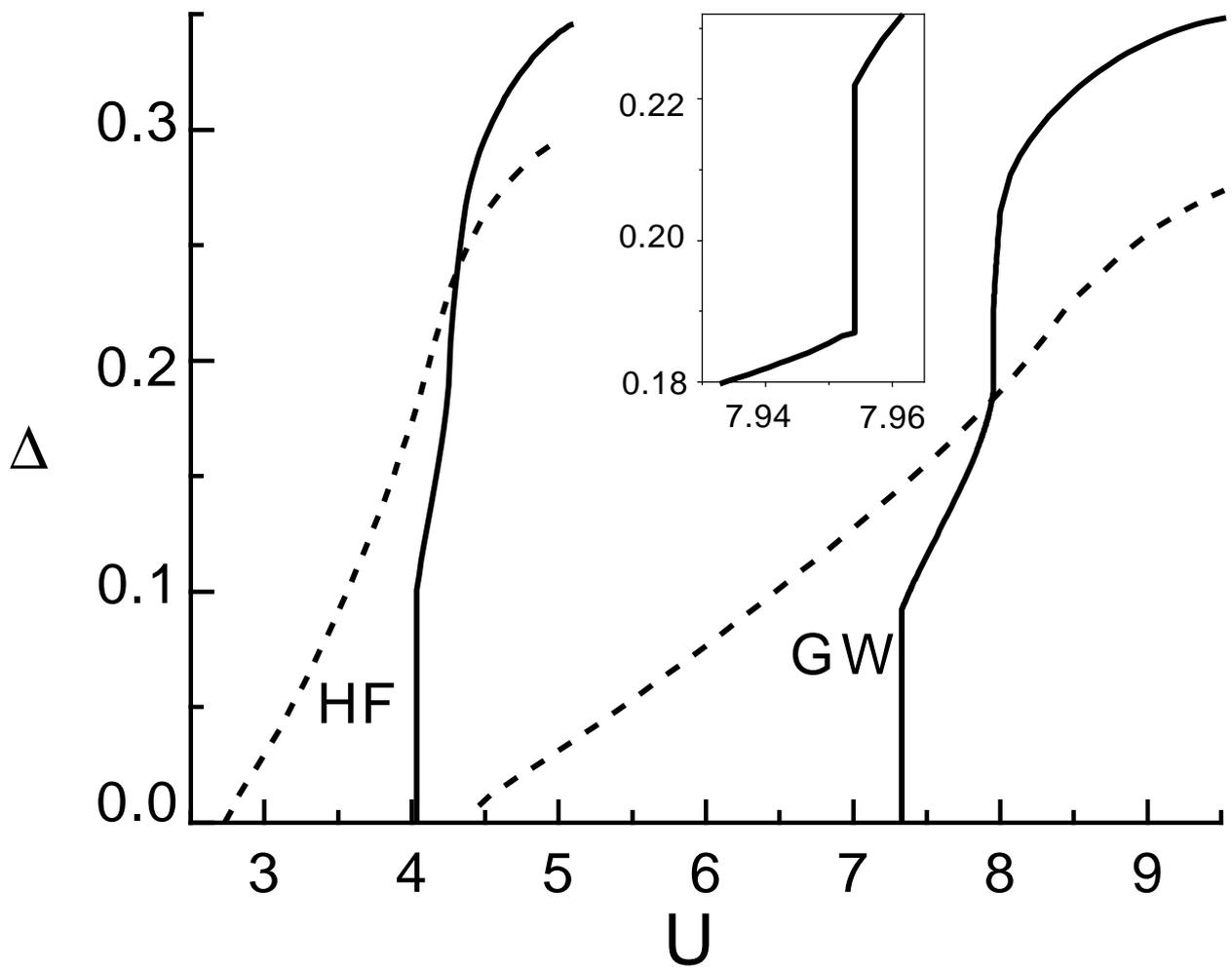
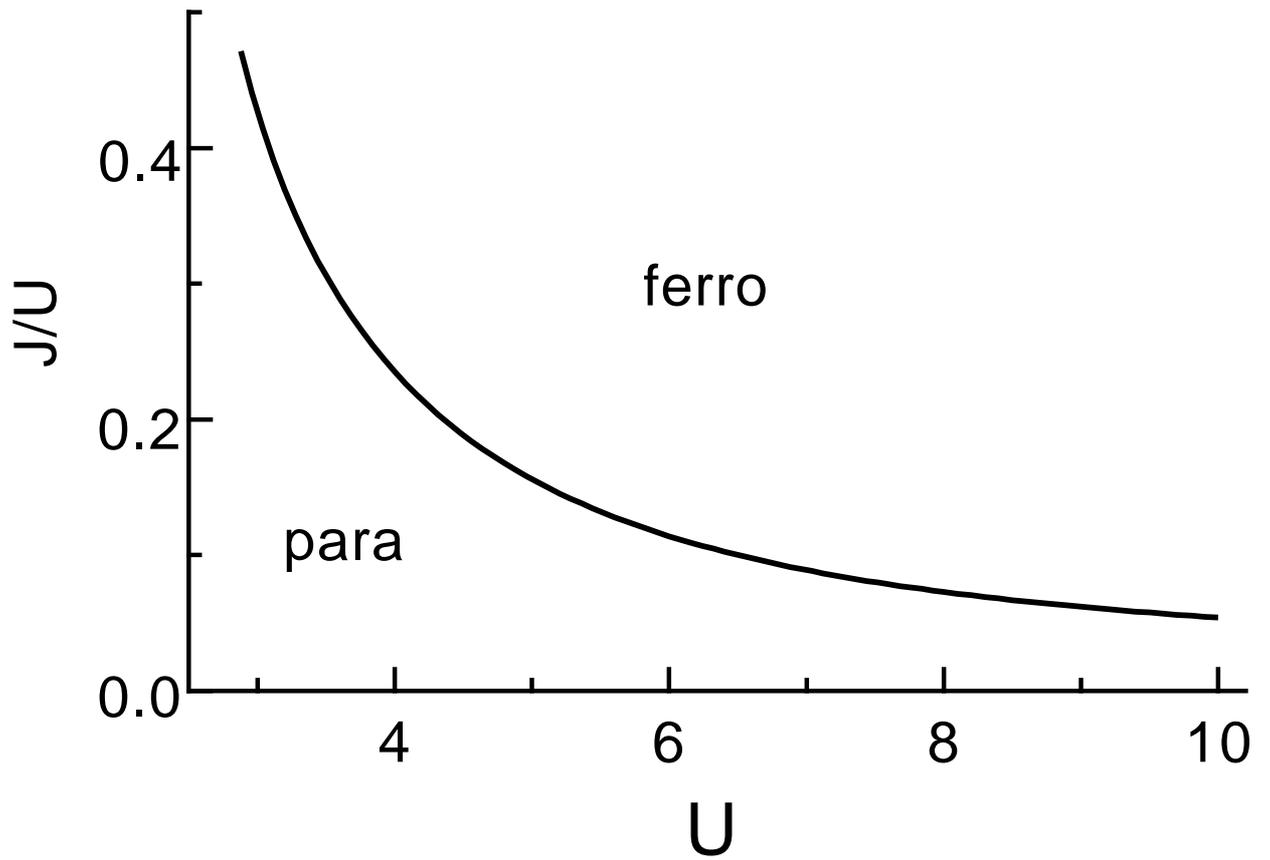
\end{document}